\documentstyle[preprint,aps]{revtex}
\begin{document}
\draft
\preprint{}
\title{Weak and Strong Localization in Low-Dimensional Semiconductor 
Structures}

\author{S.-R. Eric Yang$^{1,2}$ and J. Rammer$^{2,3}$}
\address{$^{1}$Department of Physics, Korea University, 
Seoul 136-701, Korea\\
$^{2}$Institute for Microstructural Sciences,
 National Research Council of Canada,
Ottawa, Ontario, Canada, K1A OR6\\
$^{3}$Department of Theoretical Physics,
Ume{\aa} University, 901 87 Ume{\aa}, Sweden}
\date{\today}
\maketitle

\begin{abstract}

The dependence of the localization length on the number of occupied
subbands $N$ in low-dimensional semiconductors is investigated.  
The localization length is shown to be
proportional to the number of occupied
subbands in quasi-one-dimensional quantum wires
while it grows exponentially with $N$ in quasi-two-dimensional systems.
Also a weak localization theory is developed for large N with 
a  well-defined small
expansion parameter $1/N$. The temperature dependence
of the conductivity deduced using this perturbation theory agrees
with the experimentally observed dependence.

\end{abstract}

\narrowtext

According to the one-parameter scaling theory  of 
localization \cite{G4}
electrons are localized in strictly one- and two-dimensional (1D and 2D)
systems. The applicability of the weak localization theory 
of transport \cite{Lee&Ramakrishnan} is thus limited in these systems:
it is applicable when the Thouless length, $L_T=\sqrt{D\tau_{in}}$, is
shorter than the localization length $\xi$ \cite{Th}, and when a 
small expansion parameter exists ($D$ and $\tau_{in}$ are the diffusion
constant and the inelastic scattering time). 
Moreover,  the expansion parameter of the diagrammatic               
transport theory $(k_Fl)^{1-d}$ \cite{AGD}, where $k_F$
is the Fermi wave vector, is not small for d=1, and the localization
length is comparable to the mean free path $l$.
Recent advances in nanofabrication have made it
possible to create quasi-one and -two dimensional (Q$1$D and Q$2$D) 
systems where several subbands may be occupied.  
The purpose of the present paper is to examine 
the range of the applicability of weak localization theory in 
Q1D and Q2D systems. The dependence of $\xi$ 
on the number of occupied subbands N 
determines the validity range of the condition $L_T<\xi$. 
When several subbands are occupied $(k_Fl)^{1-d}$ is no longer 
the correct expansion parameter  
and a new small expansion parameter must be found.

We apply Vollhardt and W\"{o}lfle's self-consistent diagrammatic
theory of Anderson localization \cite{V&W} to investigate these
problems. Vollhardt and W\"{o}lfle's localization length for 1D
systems is in good agreement with the exact results of Berezinskii
\cite{Berezinskii} and Abrikosov and Ryzhkin \cite{Abrikosov&Ryzhkin}.
In 2D their theory predicts that particles are always localized, in
agreement with the one-parameter scaling theory of Abrahams et al.
\cite{G4}. We extend their approach to the multisubband cases in the
present paper.  We find that in Q$1$D systems the localization length
is proportional to N while in Q$2$D systems it grows exponentially
with N. Our Q$1$D result is in good qualitative agreement with the
previous results obtained using different methods \cite{Tamura&Ando}.
The basic physics behind the dependence of the localization length on
N is the intersubband scattering which diminishes coherent
backscattering leading to longer localization lengths. We also find
that $1/N(k_Fl)^{1-d}$ serves as a small expansion parameter in Q$1$D
and Q$2$D systems. Our Q$1$D result may provide an explanation for
Mani and von Klitzing's experimentally found temperature dependence of
the conductivity \cite{Mani&vonKlitzing}($T^{-p/2}$ with $p\sim2$).
Their quantum wire has the number of occupied subbands between $7$ and
$11$, and therefore, according to our result, the localization length
may be larger than the mean free path. So for temperatures where $L_T$
is shorter than $\xi$ the conductivity may be evaluated using usual
weak localization theory.

Our interest is in conduction subbands of GaAs/AlGaAs quantum wells
and wires and other subbands with similar electronic structures.  In
Q$1$D systems only the lowest subband originating from one of the
confinement potentials is assumed to be relevant. Both in Q$1$D and
Q$2$D systems all the relevant subbands are assumed to be parabolic
with an effective mass $m^{\ast}$, so each electron state is labeled
by a subband index $n$ and a wavenumber $k$.  The level broadening due
to the elastic scattering is assumed sufficiently small in comparison
with the subband energy separations so that the subbands are well
defined.  In the presence of an external potential $V_{ext}({\bf r}) =
V_{ext}\exp\{i{\bf q}\cdot{\bf r} -i\omega t\}$ with the wave vector
${\bf q}$ parallel the to $x$-axis, the Fourier transform of the
induced density is given by $\delta n({\bf q},\omega) = \chi({\bf
  q},\omega)V_{ext}$, where the Q$2$D and Q$1$D polarization functions
are $\chi({\bf q},\omega) = \int_{}^{}\!\!dzdz'\, \chi({\bf
  q},z,z',\omega)$, and $\chi({\bf q},\omega) =
\int_{}^{}\!\!dydy'dzdz'\, \chi({\bf q},y,y',z,z',\omega)$,
respectively.  The polarization function can be expressed in terms of
the generalized diffusion coefficient $D({\bf
  q},\omega)$\cite{Forster}
\begin{equation}
\chi({\bf q},\omega) = N_F\frac{D({\bf q},\omega) q^{2}}{-i\omega + D({\bf q},\omega)q^{2}}
\end{equation}
where $N_F$ is the {\em total} $d$-dimensional density of states per 
spin at the Fermi 
energy. 
We use $\delta$-function impurity potentials 
$V_{imp} \delta({\bf r}-{\bf r_i'})$, where ${\bf r_i'}$ are the positions 
of the impurities.   
In our model, where many subbands are occupied, the Fermi energy
level broadening is assumed to be given by $\gamma=\pi N_{F} U_0$
with $U_0$ equal to $n_i|V_{imp}|^2$ ($n_i$ is the impurity concentration).

We calculate corrections to the diffusion constant by including the
weak localization effects \cite{L&K,Bergmann}. We include not only the
intrasubband, but also the intersubband scattering.  As long as the
Thouless length is shorter than the localization length the correction
to the conductivity due to weak localization effects is determined by
summing the maximally crossed diagrams \cite{L&N}. Green's functions
are specified in the subband and momentum representation by respective
set of quantum numbers.  We have for the
Cooperon
\begin{eqnarray}
\label{PPIL}
\Lambda_{m,n;m',n'}({\bf k} + {\bf k'},\omega) & = &
\Lambda_{m,n;m',n'}^0 \nonumber\\
& + & \sum_{l,l'}
\Lambda_{m,n;l,l'}^0 \Pi_{l,l'}({\bf k} + {\bf k'},\omega)
\Lambda_{l,l';m',n'}({\bf k} + {\bf k'},\omega)
\end{eqnarray}
where
\begin{equation}
\label{PI}
\Pi_{n,n'}({\bf q},\omega) = \sum_{{\bf k}}G_n^R(\epsilon +\omega,{\bf
  q} - {\bf k}) G_{n'}^A(\epsilon,{\bf k})
\end{equation}
Here 
$G_n^R(\epsilon,{\bf k}) = (\epsilon - \epsilon_{n,{\bf k}} + i/2\tau)^{-1}$
and the lifetime in each subband, $\tau=1/2\gamma$, is approximated 
to be independent of the
subband index. The subband dispersion is
$\epsilon_{n,{\bf k}} = \epsilon_{n}+{\bf k}^2/2m$ where
$\epsilon_{n}$ is the $n$'th subband energy, and
$\Lambda_{m,n;l,l'}^0$ are the
matrix elements of the impurity correlator $U_0$.

The matrix equation Eq.(\ref{PPIL}) may be solved for 
$\Lambda_{m,n;m',n'}$ perturbatively in the small frequency, long wave
length limit \cite{W}.
The matrix elements of the perturbation are
\begin{equation}
\label{matrixV}
V_{m,n;m',n'}({\bf q},\omega) = \Lambda_{m,n;m',n'}^0\left(\Pi_{n,n'}({\bf q},\omega)
- \Pi_{n,n'}({\bf 0},0)\right)
\end{equation} 
and the unperturbed eigenstates satisfy
\begin{equation}
\sum_{l,l'}\Lambda_{m,m;l,l'}^0\Pi_{l,l'}({\bf 0},0)\Phi_{l,l'}^{\alpha} = 
\lambda_{\alpha}\Phi_{m,m'}^{\alpha} \hspace{1cm} \alpha = 0,1,2,3,...
\end{equation}
where 
\begin{equation}
\label{polarization}
\Pi_{n,n'}({\bf 0},0) = \frac{2\pi iN_{n'F}}{\epsilon_{n'} -
 \epsilon_{n} +i/\tau}
\end{equation}              
A perturbation calculation yields
$\Lambda_{m,m';n,n'}({\bf q},\omega) \approx \Phi_{m,m'}^{0}\Phi_{n,n'}^{0}
/\Delta \lambda_0$  
where the change in the eigenvalue $\lambda_0$ is given by  
\begin{equation}
\label{eigenvalue}
\Delta \lambda_0 = <\!\Phi^{0}|V|\Phi^{0}\!> = \sum_{n,n'}\Phi_{n,n'}^{0} 
\Pi_{n,n'}({\bf 0},0) \left(\sum_{m,m'}V_{n,n';m,m'}({\bf k} + {\bf k'},\omega)
\Phi_{m,m'}^{0}\right)
\end{equation}              
Inserting Eq.(\ref{matrixV}) and
 Eq.(\ref{polarization}) into
Eq.(\ref{eigenvalue}) we find               
\begin{equation}
\Lambda_{m,n;m',n'}({\bf k} + {\bf k'},\omega) = \frac{-2\gamma\Phi_{m,n}^{0} 
\Phi_{m',n'}^{0}}{-i\omega + D_{eff}({\bf k} + {\bf k'})^2} \approx
\frac{-2\gamma^2/\pi N_F}
{-i\omega + D_{0}({\bf k} + {\bf k'})^2}
\end{equation}              
where the effective diffusion constant $D_{eff}$ is evaluated approximately
as follows:
Using $\Lambda_{m,m;l,l}^0=\gamma/\pi N_{F}$ and 
replacing  $\Pi_{n,n}({\bf 0},0)$ with the average 
value $<\!\Pi_{n,n}({\bf 0},0)\!>\equiv
\pi<N_{n,F}>/\gamma \approx  \pi N_F/N\gamma$, we find the eigenvector
$[\Phi_{n,n}^{0}]^2 = \gamma/\pi N_F$.
Again neglecting the off-diagonal elements of $\Pi_{n,n'}({\bf 0},0)$  
we find an approximate value for the effective 
diffusion constant
$D_{eff} \approx D_O = \sum_{n}^{}{}\raisebox{1.5ex}{$\prime$} 
D_n/N \approx <v_F>^2\tau/d$,
where the prime indicate that we only sum over occupied subbands and
$<v_F>$ is the average subband velocity on the Fermi surface.
In our approximation each occupied subband has the same Cooperon
and contribute to the 
conductivity independently.  The weak localization correction leads then 
to a conductivity  per spin of a Q$1$D system
\begin{equation}
\label{WLcorrection2}
\sigma = \sigma_0 + \delta\sigma = \frac{<\!n\!>Ne^2\tau}{m}
- \frac{e^2}{\hbar\pi}(L-l) = 
\sigma_0[1-\frac{1}{N}\left(\frac{L}{l}-1\right)]
\end{equation}
The average density per subband $<\!n\!>$ is $<\!k_F\!>/\pi$. From 
Eq.(\ref{WLcorrection2}) we see
that the correction due to weak localization effects is of order $1/N$. 
When several subbands are occupied in Q1D and Q2D systems
$1/N(k_F\ell)^{d-1}$ serves as the small expansion parameter.  

We now proceed to derive a formally exact expression for the inverse
diffusion constant. We will then find an approximate expression for
the inverse diffusion constant by replacing the irreducible vertex by
the lowest order expression. The resulting equation is then extended
into the insulator regime by demanding self-consistency.  Our
derivation is an extension of Vollhardt and W\"{o}lfle's method to QD
systems.

The impurity average of the product of Green's functions $G_{m{\bf k}_{+}}^R
G_{m'{\bf k'}_{+}}^A$ may be expressed as
\begin{equation}
  \Phi_{m,n,{\bf k};m',n',{\bf k'}}({\bf q},\omega) = G_{m{\bf
      k}_{+}}^R G_{n{\bf k}_{-}}^A \delta_{{\bf k},{\bf
      k'}}\delta_{n,n'}\delta_{m,m'} + G_{m{\bf k}_{+}}^R G_{n{\bf
      k}_{-}}^A\Gamma_{m,n,{\bf k};m',n',{\bf k'}}({\bf q},\omega)
  G_{m'{\bf k'}_{+}}^R G_{n'{\bf k'}_{-}}^A
\end{equation}
The vertex function
$\Gamma_{m,n,{\bf k};m',n'{\bf k'}}({\bf q},\omega)$ is given in terms 
of the irreducible
vertex function $U_{m,n,{\bf k};m',n',{\bf k'}}({\bf q},\omega)$
\begin{eqnarray}
\label{aa}
& & \Gamma_{m,n,{\bf k};m',n',{\bf k'}}({\bf q},\omega)  = 
U_{m,n,{\bf k};m',n',{\bf k'}}({\bf q},\omega) \nonumber\\ 
& + &  \sum_{m'',n'',{\bf k''}} 
U_{m,n,{\bf k};m'',n'',{\bf k''}}({\bf q},\omega) G_{m''{\bf k''}_{+}}^R
G_{n''{\bf k''}_{-}}^A \Gamma_{m'',n'',{\bf k''};m',n',{\bf k'}}
({\bf q},\omega)
\end{eqnarray}
Using Eq.(\ref{aa}) $\Phi$ may be written in 
terms of the irreducible vertex function
\begin{equation}
\label{Phivertex}
\Phi_{n,{\bf k};n',{\bf k'}}({\bf q},\omega) = G_{n{\bf k}_{+}}^R
G_{n{\bf k}_{-}}^A \delta_{{\bf k},{\bf k'}}\delta_{n,n'} + G_{n{\bf
    k}_{+}}^R G_{n{\bf k}_{-}}^A \sum_{n'',{\bf k''}} U_{n,{\bf
    k};n'',{\bf k''}}({\bf q},\omega) \Phi_{n'',{\bf k''};n',{\bf
    k'}}({\bf q},\omega)
\end{equation}
Using
\begin{equation}
G_{n{\bf k}_+}^RG_{n{\bf k}_-}^A  = \frac{\Delta G_{n,{\bf k}}}{\omega -
{\bf k}\cdot{\bf q} - \Delta \Sigma_{n,{\bf k}}}
\end{equation}
we rewrite Eq.(\ref{Phivertex})
\begin{eqnarray}
\label{Phivertex2}
(\omega - {\bf k}\cdot{\bf q} - \Delta \Sigma_{n,{\bf k}})\Phi_{n,{\bf
    k};n',{\bf k'}}({\bf q},\omega) & = & -\Delta G_{n,{\bf k}} \Bigg(
\delta_{{\bf k},{\bf k'}}\delta_{n,n'} \nonumber\\ & + &
\sum_{n'',{\bf k''}} U_{n,{\bf k};n'',{\bf k''}}({\bf q},\omega)
\Phi_{n'',{\bf k''};n',{\bf k'}}({\bf q},\omega) \Bigg)
\end{eqnarray}
Summing over $n,{\bf k};n',{\bf k'}$ in Eq.(\ref{Phivertex2}), and using 
$\Delta G_{n,{\bf k}}=2i\Im mG_{n,{\bf k}}$, and the Ward identity
\begin{equation}
  \Delta \Sigma_{n,{\bf k}} = \sum_{n',{\bf k'}} U_{n,{\bf k};n',{\bf
      k'}}({\bf q},\omega) \Delta G_{n',{\bf k'}}
\end{equation}
we find 
\begin{equation}
\label{phieq}
\omega\Phi({\bf q},\omega) - q\Phi_j({\bf q},\omega) =  2\pi iN_F
\end{equation}
where $\Phi({\bf q},\omega) = \sum_{n,{\bf k};n',{\bf k'}}
\Phi_{n,{\bf k};n',{\bf k'}}({\bf q},\omega)$ and $\Phi_j({\bf
  q},\omega) = \sum_{n,{\bf k};n',{\bf k'}}\frac{{\bf k}
  \cdot\hat{{\bf q}}}{m^*} \Phi_{n,{\bf k};n',{\bf k'}}({\bf
  q},\omega)$.

Multiplying Eq.(\ref{Phivertex2}) with ${\bf k}\cdot\hat{{\bf q}}/k_{nF}$,
where $k_{nF}$ is the n'th subband Fermi wave vector, and summing 
over $n,{\bf k};n',{\bf k'}$ we find when $\gamma>>\omega$ 
\begin{equation}
\label{kineq}
K({\bf q},\omega)\Phi_j({\bf q},\omega)  + iqD_0\Phi({\bf q},\omega) = 0
\end{equation}
where 
\begin{equation}
\label{Kexp}
K(\omega,{\bf q}) = 1 + \tau\sum_{n,{\bf k};n',{\bf k'}} \frac{d}{\pi
  k_{nF}^2N_{nF}N} {\bf k}\cdot\hat{{\bf q}} \Delta G_{n,{\bf k}}
\left(U_{n,n;n',n'}({\bf k} + {\bf k'},\omega) - U_0\right) \Delta
  G_{n',{\bf k'}} {\bf k'}\cdot\hat{{\bf q}}
\end{equation}
In deriving Eq.(\ref{kineq}) we have used the approximation $\Delta
\Sigma_{n,{\bf k}} = \Sigma_{n,{\bf k}}^R(E_F+\omega) - \Sigma_{n,{\bf
    k}}^A(E_F) \approx 2i\Im m \Sigma_{n,{\bf k}}^R(E_F) = 2i\gamma$,
as $\omega<<E_F$.  In addition we have used the peaked character of
the spectral function to approximate
\begin{eqnarray}
  \sum_{n',{\bf k'}} \Phi_{n,{\bf k};n',{\bf k'}}({\bf q},\omega) &
  \approx & \frac{\Delta G_{n,{\bf k}}}{-2\pi iN_{nF}N}
  \left(\sum_{n',{\bf k'};n'',{\bf k''}} \Phi_{n',{\bf k'};n'',{\bf
        k''}}({\bf q},\omega) \right. \nonumber\\ & + & \left.
    \frac{{d\bf k}\cdot\hat{{\bf q}}}{k_{nF}} \sum_{n',{\bf
        k'}}\frac{{\bf k'}\cdot\hat{{\bf
          q}}}{k_{nF}^{'}}\sum_{n'',{\bf k''}} \Phi_{n',{\bf
        k'};n''{\bf k''}}({\bf q},\omega) \right)
\end{eqnarray}
keeping the first two terms in the Legendre expansion.  Here we note
that ${\bf k}\cdot\hat{{\bf q}}= k_{nF}P_l(\cos\theta)$, where
$\theta$ is the angle between ${\bf k}$ and $\hat{{\bf q}}$.

Using the general form of the response kernel \cite{V&W} and 
approximating $U_{n,n;n,n} - U_0$ by $\Lambda_{n,n;n,n}$ we find
\begin{equation}
\label{Domrgaeq}
\frac{D_0}{D({\bf q},\omega)}  =  1 + \frac{1}{\pi N_F}\sum_{{\bf Q}}
\frac{1}{-i\omega + D_0{\bf Q}^2}
\end{equation}
The residue of $\Lambda_{n,n,n,n}$ at $i\omega=D_0q^2$ is $2\gamma/\pi
N_F$ and is about $1/N$ times smaller than the value for the case
where only one subband is occupied.  This will have important
consequences for the subband dependence of the localization length.
Invoking self-consistency and multiplying Eq.(\ref{Domrgaeq}) with
$D(\omega)/D_0$ we find
\begin{equation}
\label{Domrgaeq2}
\frac{D(\omega)}{D_0}  =  1 - \frac{1}{\pi N_FD_0}\sum_{{\bf Q}}
\frac{1}{-i\omega/D(\omega) + {\bf Q}^2}
\end{equation}

This result determines the dependence 
of the localization length on the number of occupied subbands.
By setting the upper $Q$-integration limit in Eq.(\ref{Domrgaeq2}) 
to $1/l$ and introducing a
new variable $y=\sqrt{D(\omega)/-i\omega} Q$ we find
\begin{equation} 
\label{Domrgaeq3}
\frac{D(\omega)}{D_0}  =  1 - \frac{2\pi^{d-1}l^{2-d}}{\pi N_FD_0(2\pi)^d}
\tilde{\xi}^{2-d}
\int_{0}^{\tilde{\xi}}\!\!\!dy\, \frac{y^{d-1}}{1+y^2}
\end{equation}
where $\tilde{\xi} = \sqrt{D(\omega)/-i\omega l^2} = \xi/l$.  In the
insulating regime the generalized diffusion constant approaches zero
as $\omega$ approaches zero \cite{Gtze}, and we find for Q1D and Q2D
systems $\tilde{\xi}\tan^{-1}\tilde{\xi} = N\pi$ and $\xi =
l\left(e^{N/\lambda} - 1\right)^{1/2}$ where $\lambda =1/2\pi
E_F\tau$.  For large $N$, we thus have $\xi = 2Nl$ for Q1D systems.
When $N=1$ and $E_F>>\tau$ both results for Q$1$D and Q$2$D agree well
with the perturbative estimates of $\xi$ in 1D and 2D systems, namely
$\pi l$ and $le^{\pi k_Fl/2}$.  We thus predict for quasi-dimensional
systems a cross-over from the strongly localized to the weakly
localized regime as the number of occupied subbands increases.  The
fact that the residue of $\Lambda_{n,n;n,n}$ is 1/N times smaller than
that of singly occupied subbands is responsible for this effect.  It
is simple to show that the scaling functions $\beta$ of Q1D and Q2D
systems are not changed from the corresponding 1D and 2D values.

For sufficiently large N the condition $L_T<\xi$ can be fulfilled and
weak localization theory may be used to calculate the conductivity.
We find in Q$1$D 
systems that $\delta \sigma\sim T^{-1}$.   The effect
of Coulomb interactions in the presence of weak localization effects
in 1D and 2D systems  has been investigated by Al'tshuler 
et al \cite{Al'tshuler}.  In Q1D and Q2D systems the same 
diagrams of Al'tshuler et al may be used when the Cooperon is replaced 
by our expression.  
The temperature dependence of these corrections is  
$T^{-1/2}$ in Q1D systems.   The temperature exponents found above, 
-1 and -1/2, are in agreement
with the recent experimental values \cite{Mani&vonKlitzing}.
Mani and von Klitzing observed that the magnetoresistance
of  GaAs/AlGaAs
quantum wires exhibits
$T^{-1}$ and $T^{-1/2}$ behavior above 1.5 K, and
at temperatures below 1.5 K the observed magnetoresistance
saturates, as expected from the effect of electron localization.  
The number of occupied subbands in
one of the wires used in the experiment
is estimated to be between 7 and 11 \cite{Mani&vonKlitzing2}. 
We believe that this large number for the occupied subbands is responsible
for the observed temperature exponents.

This work was supported by Korea University through a start-up
grant and by NON DIRECTED RESEARCH FUND, Korea Research Foundation. 
Also we acknowledge the support by
the Swedish Natural Research Council through contract F-AA/FU $10199-306$.

\end{document}